# High-efficiency integrated laser on erbium-doped lithium niobate-on-insulator

Chunyu Zhang[1#], Yuqi Zhang[1#], Yiyang Zou[1#], Xiaomin Wang[1], Binwen Niu[1], Cangsu Yuan[1], Tongxin Xue[1], Chunlin Zhu[1], Hongde Liu[1✉], Dahuai Zheng[1✉], Shiguo Liu[1], Fang Bo[1✉], Yongfa Kong[1✉] and Jingjun Xu[1✉]

[1]MOE Key Laboratory of Weak-Light Nonlinear Photonics, School of Physics and TEDA Institute of Applied Physics, Nankai University, Tianjin 300071, China

[#] These authors contribute equally to this work

## Abstract

Lithium niobate on insulator (LNOI) combines the outstanding optical properties of lithium niobate (LN) with strong optical confinement, scalable fabrication and high-density integration, making it a leading platform for integrated photonic chips. Recent advances in LNOI photonics have mainly centred on passive and electro-optic components, including couplers, waveguides, microcavities and modulators, whereas efficient on-chip laser sources remain insufficiently developed, limiting the realization of fully integrated LN photonic systems. Because LN is an indirect-bandgap material, lasing on LNOI generally relies on photoluminescence from rare-earth-ion doping, yet the conversion efficiency of doped LNOI lasers has remained low. By comparing LNOI microcavity lasers with fibre lasers and waveguide amplifiers, we identify the limited number of rare-earth ions participating in stimulated emission as a key factor responsible for inefficient pump utilization. Here we demonstrate an integrated Er-doped LNOI laser that combines high-quality, highly Er-doped LN, a large-diameter wide-microring resonator, a low-loss waveguide amplifier and bidirectional pumping. This architecture enables a slope efficiency of 16.91% at 1562 nm, exceeding 10% on the LNOI platform for the first time. Our results provide a route towards high-efficiency LNOI lasers for fully integrated photonic systems.

The rapid growth of artificial intelligence, big data and the Internet of Things has created an increasing demand for high-capacity information processing[1]. However, as electronic devices continue to scale down, integrated electronic chips are approaching fundamental physical limits[2-4]. Integrated photonics offers a promising route towards next-generation information technologies by enabling high-speed, low-loss and energy-efficient signal processing[5]. Among the various material platforms, LN is particularly attractive owing to its broad optical transparency window, large electro-optic coefficient and strong nonlinear optical response, while LNOI further combines these intrinsic material advantages with strong optical confinement and compatibility with scalable nanofabrication processes[6,7]. Over the past decade, LN-based micro- and nanophotonic devices have evolved into a comprehensive functional library, including directional couplers[8], multimode interferometers[9], Mach-Zehnder interferometers[10], optical resonators[11], grating couplers[12,13], optical parametric amplifiers[14,15] and high-performance electro-optic modulators[16-18], thereby laying the foundation for large-scale integrated photonic systems[19]. Nevertheless, despite substantial progress in on-chip light routing, resonant control, nonlinear frequency conversion and electro-optic modulation, efficient on-chip light sources remain a critical missing element for fully integrated LNOI photonic circuits. Because LN is an indirect-bandgap material, it is intrinsically inefficient for electroluminescence[20], and laser emission on the LNOI platform generally relies on photoluminescence from rare earth ion doping[21]. Up to now, the conversion efficiency of doped LNOI lasers remains very low, which severely restricts the development of LN integrated photonics.

Since 2021, extensive efforts have been devoted to improving the slope efficiency of LNOI lasers[22-40]. Device parameters and architectures, including microcavity design, fabrication processes, coupling schemes and heterogeneous or monolithic integration, have been systematically investigated. These studies have increased the slope efficiency of Er:LNOI lasers from the initial level of $10^{-5}$% to $10^{-1}$%[22-33]. Nevertheless, further progress has reached a bottleneck since 2023, with the highest reported slope efficiencies remaining at the $10^{-1}$% level[34,36,37]; only one report has exceeded 1%, reaching the current maximum value of 2.51%[38]. Therefore, breaking through the efficiency limitation of Er:LNOI lasers has become an urgent challenge. Addressing this challenge may require a broader perspective beyond the conventional framework of Er:LNOI microcavity lasers. For comparison, the slope efficiency of fibre lasers has exceeded 80%[41-43]. This large performance gap suggests that the population of Er ions participating in stimulated emission in Er:LNOI microcavities is far smaller than that in fibre lasers. Consequently, the pump light confined in the microcavity cannot be efficiently converted into laser emission and is instead largely dissipated through propagation loss and residual absorption. Therefore, improving the slope efficiency of Er:LNOI lasers should start with increasing the number of optically active Er ions inside the microcavity. On the other hand, Er:LNOI waveguide amplifiers provide an important reference: Er-doped waveguide amplifiers can already achieve efficient optical signal amplification[44], with recent reports reporting net gains exceeding 30 dB and the optical conversion efficiency substantially higher than that of Er:LNOI lasers[45,46]. Integrating an Er:LNOI laser with an optical amplifier should therefore markedly enhance the slope efficiency of the integrated laser by involving a much larger number of Er ions in stimulated emission. Moreover, previous studies have shown that backward pumping, and especially bidirectional pumping, can significantly improve the net gain of Er:LNOI waveguide amplifiers[47].

Guided by the above analysis, we designed an integrated Er:LNOI laser that simultaneously increases the number of Er ions involved in stimulated emission and enhances pump-light utilization. To increase the population of luminescent Er ions in the microcavity, we grew a high-quality LN crystal doped with 1.0 mol% Er and fabricated a large-area microring resonator with a radius of 200 μm and a waveguide width of 40 μm. For the integrated amplifier, three low-loss spiral waveguide amplifiers were cascaded to provide a total gain length of 4.43 cm, together with a bidirectional pumping. Benefiting from this combined resonator-amplifier architecture, the integrated Er:LNOI laser achieves a slope efficiency of 16.91% at 1562 nm, exceeding 10% on the LNOI platform for the first time. Our results establish a design strategy for high-efficiency Er:LNOI lasers and provide important guidance for the development of high-performance on-chip LN light sources.

# Results

## Fabrication of Er:LNOI and On-Chip Structures

To improve the slope efficiency of Er:LNOI laser, 3-inch 1.0 mol% Er:LN crystals were grown (Fig. 1a). Although it was reported that the optimal doping concentration of Er ions in Er:LN crystals is about 2.0 mol%[48], the growth of large-diameter single crystals at such a high doping level remains challenging. More importantly, it is essential to ensure the optical quality, which must be maintained to support low-loss integrated photonic devices. The grown crystal was sliced into 3-inch wafers (Fig. 1b). X-ray rocking-curve measurements gave a Lorentzian-fitted FWHM of 64 arcsec, confirming high crystalline quality (Fig. 1c). Figure 1d illustrates the fabrication process of the on-chip structures, including Er:LNOI wafer preparation, high-precision electron-beam lithography (EBL), inductively coupled plasma reactive-ion etching (ICP-RIE) and buffered oxide etching (BOE). The final stack comprised a 500-μm Si handle, a 2.0-μm $SiO_2$ layer, and a 600-nm Er:LN device layer, which was etched to 350 nm by ICP-RIE for photonic structure definition.

To balance cost, practicality, and scalable integration, we designed a wide-microring resonator architecture and integrated it with a spiral waveguide amplifier (Fig. 1e). The three structures shown from top to bottom are the standalone wide-microring resonator, the integrated resonator-amplifier structure consists of the same wide-microring resonator and waveguide amplifier, and the standalone waveguide amplifier. Figure 1f shows an optical microscope image of the fabricated integrated device, and Fig. 1g presents a magnified view of the spiral amplifier region. The fabricated resonator and waveguide amplifier exhibit well-defined patterns and smooth morphology, confirming fabrication reliability.

For the wide-microring resonator, the outer and inner radii are 200 μm and 160 μm, respectively. The resonator is coupled to a ridge bus waveguide for optical input and output. To support single-mode operation, the ridge waveguide was designed with a width of 1.2 μm, and the gap between its upper edge and the wide-microring resonator was set to 0.6 μm. The spiral waveguide amplifier consists of three identical spiral ridge waveguides, each with a width of 1.2 μm. The innermost spiral has a radius of 100 μm, with an inter-spiral spacing of 2 μm, as shown in Fig. 1g. The three cascaded spiral waveguides provide a total gain length of 4.43 cm within a compact footprint of 1.9 mm × 0.45 mm, balancing optical gain enhancement with integration density.

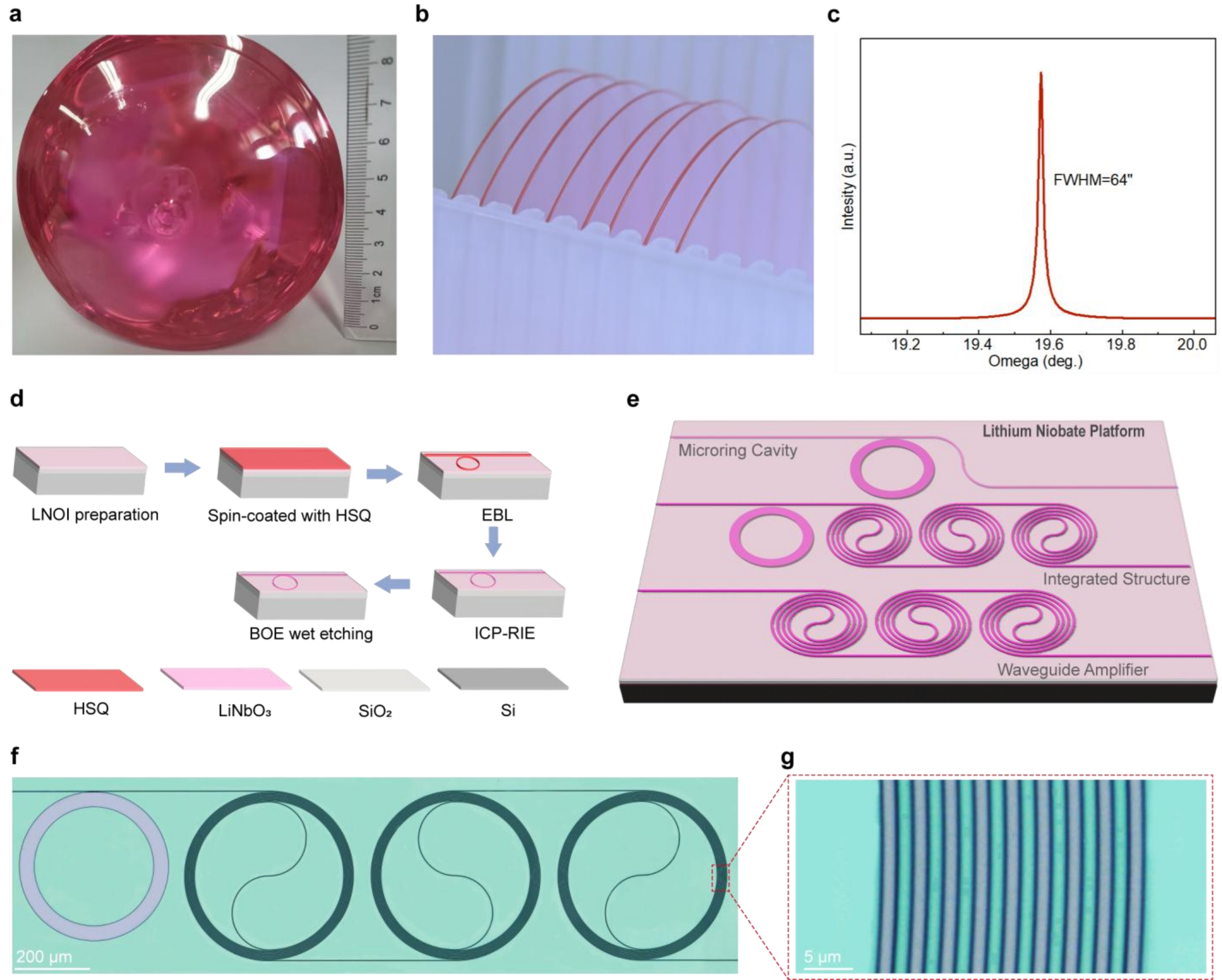

**Fig. 1. Fabrication process of on-chip Er:LNOI lasers and amplifiers.** (**a**) 3-inch Er:LN crystal. (**b**) 3-inch Er:LN wafer. (**c**) Rocking curve of the Er:LN wafer. (**d**) Process flow for on-chip structure fabrication. (**e**) Schematic illustrations of the fabricated standalone wide microring resonator, the integrated structure of the standalone wide microring resonator with amplifier, and standalone waveguide amplifier. (**f**) Optical micrograph of the integrated Er:LNOI wide microring cavity and waveguide amplifier structure. (**g**) Magnified optical micrograph of the amplifier region in (f).

## Characterization of the Wide Microring Er:LNOI Laser

To increase the number of Er ions participating in the gain process, a large-radius microring resonator was fabricated. As shown in Fig. 2a, the optical setup was used for characterization. A microring radius of 200 μm was selected, which is substantially larger than the previously reported radii of 40 μm and 100 μm[22,24]. A larger microcavity, such as a resonator with a radius of 500 μm[38], was not adopted because of its limited compatibility with dense on-chip integration.

Although microdisk resonators are generally thought to involve more Er ions, the optical modes in both microdisks and microrings are mainly confined near the edge, suggesting that a sufficiently wide microring can provide comparable ion utilization with a smaller lithography area. To explore this trade-off, we fabricated and tested microring resonators with different ring widths (all 200 μm radius) alongside microdisk resonators under 1486 nm pumping. Based on the measured performance, a ring width of 40 μm was selected because it gave the highest emission efficiency among the microrings and was comparable to the microdisk (other geometries are shown in Supplementary Fig. S1). Therefore, a wide-microring resonator with an outer radius of 200 μm and a width of 40 μm was adopted as the optimized configuration, balancing increased Er ions participation with compact on-chip integration.

Figure 2b shows the broadband transmission spectrum of the fabricated wide-microring resonator, from which a free spectral range (FSR) of 0.85 nm is extracted. The loaded quality factor ($Q_L$), obtained by Lorentzian fitting of the resonance centred near 1562.72 nm, is $1.67 \times 10^5$, as shown in Fig. 2c. Utilizing the measured FSR and $Q_L$, the intrinsic quality factor ($Q_0$) is calculated to be $2.85 \times 10^5$, corresponding to a waveguide propagation loss of 1.40 dB $cm^{-1}$ near 1560 nm. These experimentally extracted parameters were then used as key inputs for the subsequent propagation-loss analysis and amplifier simulations.

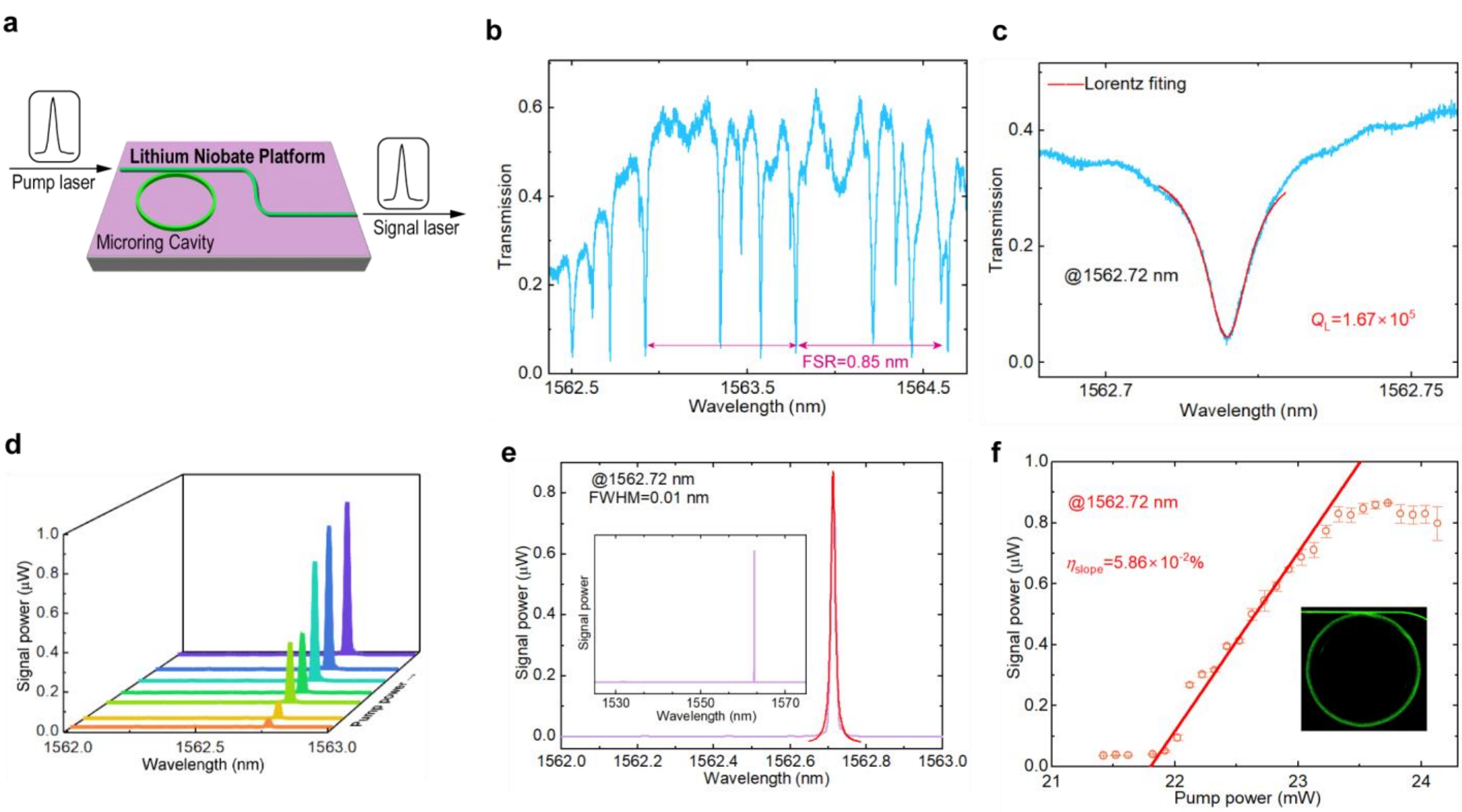

**Fig. 2. Characteristics of the standalone wide microring Er:LNOI laser.** (**a**) Optical path schematic of the standalone wide microring resonator. (**b**) Transmission spectrum of 1560 nm measured by an oscilloscope. (**c**) Transmission spectrum near 1562.72 nm and its Lorentzian fitting, yielding a $Q_L$ factor of $1.67\times10^5$. (**d**) On-chip lasing spectra corresponding to different on-chip pump powers. (**e**) Broadband (inset) and narrowband wavelength scans of the on-chip output laser spectrum measured by an optical spectrum analyser. (**f**) On-chip output laser power versus pump power, showing a slope efficiency of $5.86\times10^{-2}$%.

The evolution of the emission spectra with increasing pump power is shown in Fig. 2d. Representative laser spectrum recorded using an optical spectrum analyser (OSA) is presented in Fig. 2e. The inset shows a broadband spectrum over the 1530-1570 nm wavelength range, where a single dominant lasing peak is observed. Lorentzian fitting of the high-resolution spectrum gives a full width at half maximum (FWHM) of 0.01 nm. The output-power characteristics are summarized in Fig. 2f. Linear fitting of the power-transfer curve yields a lasing threshold of 21.80 mW and a slope efficiency of $5.86 \times 10^{-2}$%. The maximum output power reaches 0.86 μW at a pump power of 23.73 mW. This slope efficiency is comparable to those of previously reported Er:LN microcavity lasers[22-35].

## Characterization of the Amplifier Integrated Er:LNOI Laser

Although standalone Er:LNOI microring lasers have demonstrated continuous performance improvements[22-35], their output power and efficiency remain constrained by the limited gain volume within the resonator. The above experimental results indicate that simply increasing the Er ions concentration in the microring cannot break the bottleneck of the laser's slope efficiency. Conversely, Er:LNOI waveguide amplifiers have achieved net gains exceeding 30 dB[49], emphasizing the potential of amplifier-assisted laser architectures. Inspired by this, we adopted an integrated approach combining a laser resonator with a waveguide amplifier. After characterizing the standalone wide-microring laser, we focused on designing and optimizing the amplifier prior to integration.

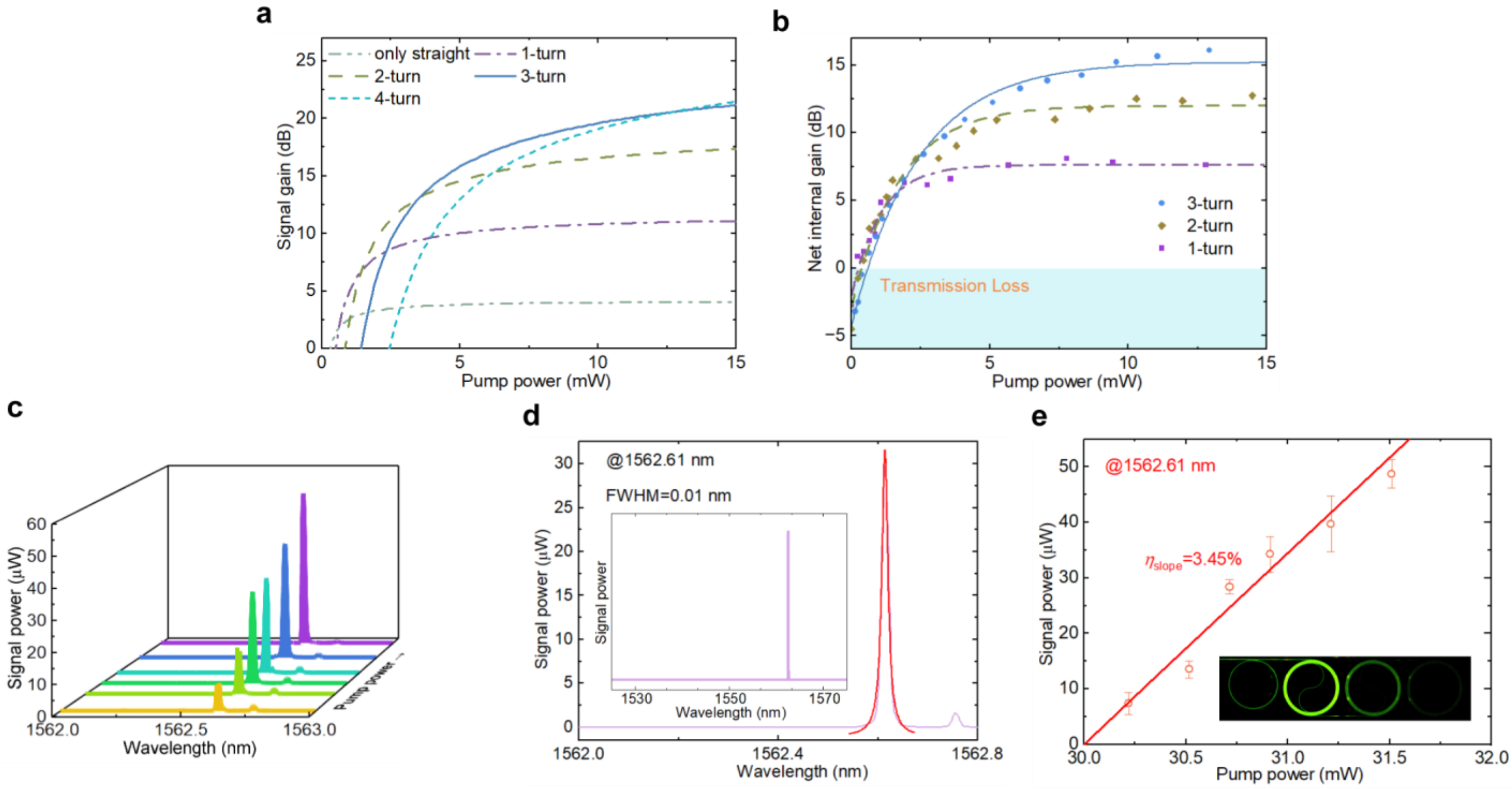

**Fig. 3. Characteristics of Er:LNOI amplifier and the integrated Er:LNOI laser with amplifier.** (**a**) Simulation results as a function of spiral waveguide amplifiers with different numbers. (**b**) Experimental results of spiral waveguide amplifiers with different numbers. (**c**) On-chip lasing spectra corresponding to different on-chip pump powers. (**d**) Broadband and narrowband wavelength scans of the on-chip output laser spectrum measured by an optical spectrum analyser. (**e**) On-chip output laser power versus pump power, showing a slope efficiency of 3.45%.

We used a spiral waveguide (width 1.2 μm for single-mode, Fig. 3a) to maximize gain with a compact footprint. Numerical simulations were conducted for amplifiers with varying numbers of spiral turns; the underlying models and parameters are provided in Eqs. (S1)-(S9) and Fig. S3 of the Supplementary Information. Among the configurations examined, the 3-turn amplifier exhibited the highest gain at low and moderate pump powers, whereas the 4-turn design experienced reduced gain due to increased propagation loss and non-uniform pump absorption. At higher pump powers, the performance gap between the two configurations diminished. Considering both gain performance and device footprint, the 3-turn geometry was identified as the most suitable for integration. To validate the simulations, 1-3 turn amplifiers were fabricated and tested. As shown in Fig. 3b, measured gains closely matched the numerical predictions, with the 3-turn amplifier achieving a maximum gain of 15.16 dB. This 3-turn amplifier was therefore integrated with the wide-microring laser.

Experimental results of the integrated Er:LNOI laser are presented in Figs. 3c-e. Figure 3c displays the emission spectra under different pump powers. The inset of Fig. 3d presents a full spectral scan across the C-band acquired using an OSA. The main panel shows the high-resolution spectrum around the emission peak at 1562.61 nm. Lorentzian fitting yields a narrow FWHM of 0.01 nm, consistent with the standalone microring resonator and confirming excellent monochromaticity. For the integrated structure, a slope efficiency of 3.45% was obtained around the emission wavelength of 1562.61 nm, placing it among the internationally leading performance levels. Under a pump power of 31.51 mW, the maximum output power reached 48.71 μW. The threshold power, determined via linear fitting, was calculated to be 30.00 mW, as shown in Fig. 3e.

By comparing the integrated laser in Fig. 3e with the standalone microring laser in Fig. 2f, we find that the slope efficiency increases from $5.86 \times 10^{-2}$% to 3.45%, corresponding to an enhancement of approximately 60-fold, or 17.70 dB.

The maximum output power also increases from 0.86 μW for the standalone microring laser to 48.71 μW for the integrated laser, showing a similar enhancement trend. This improvement is consistent with the gain-enhancement effect of the 3-turn waveguide amplifier, which provides a maximum gain of 15.16 dB. These results indicate that the substantial increases in both slope efficiency and output power mainly originate from the integrated amplifier, confirming the effectiveness of the resonator-amplifier architecture for improving the performance of Er:LNOI lasers.

## Characterization of the Integrated Er:LNOI Laser under Bidirectional Pumping

These results highlight the importance of gain enhancement for high-performance laser operation. Previous studies have shown that backward pumping, and especially bidirectional pumping, can significantly increase the net gain of waveguide amplifiers[47]. We further characterized the integrated Er:LNOI laser under bidirectional pumping, with careful optimization of the pump-power ratio between the forward and backward directions.

Figure 4a shows the schematic of the optical path used for bidirectional pumping testing of the integrated laser. A fixed-wavelength 1486-nm laser was used, with a beam splitter and an attenuator to adjust the forward and backward pump intensities, and the laser signal was finally measured by an OSA. Numerical simulations were first performed to evaluate the gain characteristics of the 3-turn amplifier under different pumping schemes (Fig. 4b). Bidirectional pumping provides the highest gain, outperforming both backward and forward pumping configurations. This behaviour was corroborated experimentally, as shown in Fig. 4c. For the 3-turn waveguide amplifier, we also characterized its performance under backward pumping; the results are shown in Fig. 4d, where the gain gradually decreases as the signal power increases, consistent with the saturation behaviour of Er-doped amplifiers.

The performance of the integrated laser was experimentally characterized under bidirectional pumping with a forward-to-backward pump power ratio of 15.8, using the setup shown in Fig. 4a. Figure 4e presents the spectra scanned by an OSA under different pump powers, demonstrating relatively stable central wavelength operation. In Fig. 4f, the inset shows the emission profile across the C-band scanned by the OSA, where only a single dominant peak is observed. The main panel displays a detailed view of this individual emission peak. The results of bidirectional pumping are presented in Fig. 4g. A remarkable slope efficiency of 16.91% was achieved, representing the first result above 10% for rare-earth ions doped LNOI lasers. Under a pump power of 25.29 mW, this configuration attained a maximum output power of 43.12 μW. The lasing threshold was calculated to be 25.01 mW. These results establish bidirectional pumping as an effective approach to increasing the slope efficiency of integrated Er:LNOI laser.

Bidirectional pumping with different ratios of forward to backward pump power was examined. Figure 4h shows the experimental results for bidirectional pump ratios ranging from 6 to 25. It can be seen that an optimal pump ratio exists within this range, which corresponds to our experimental results shown in Fig. 4g. And in Fig. 4h, three experimental pump ratios yield slope efficiencies exceeding 10% (Figs. 4g, S2), this indicates that our integrated Er:LNOI laser achieves high slope efficiencies over a wide range of pump ratios.

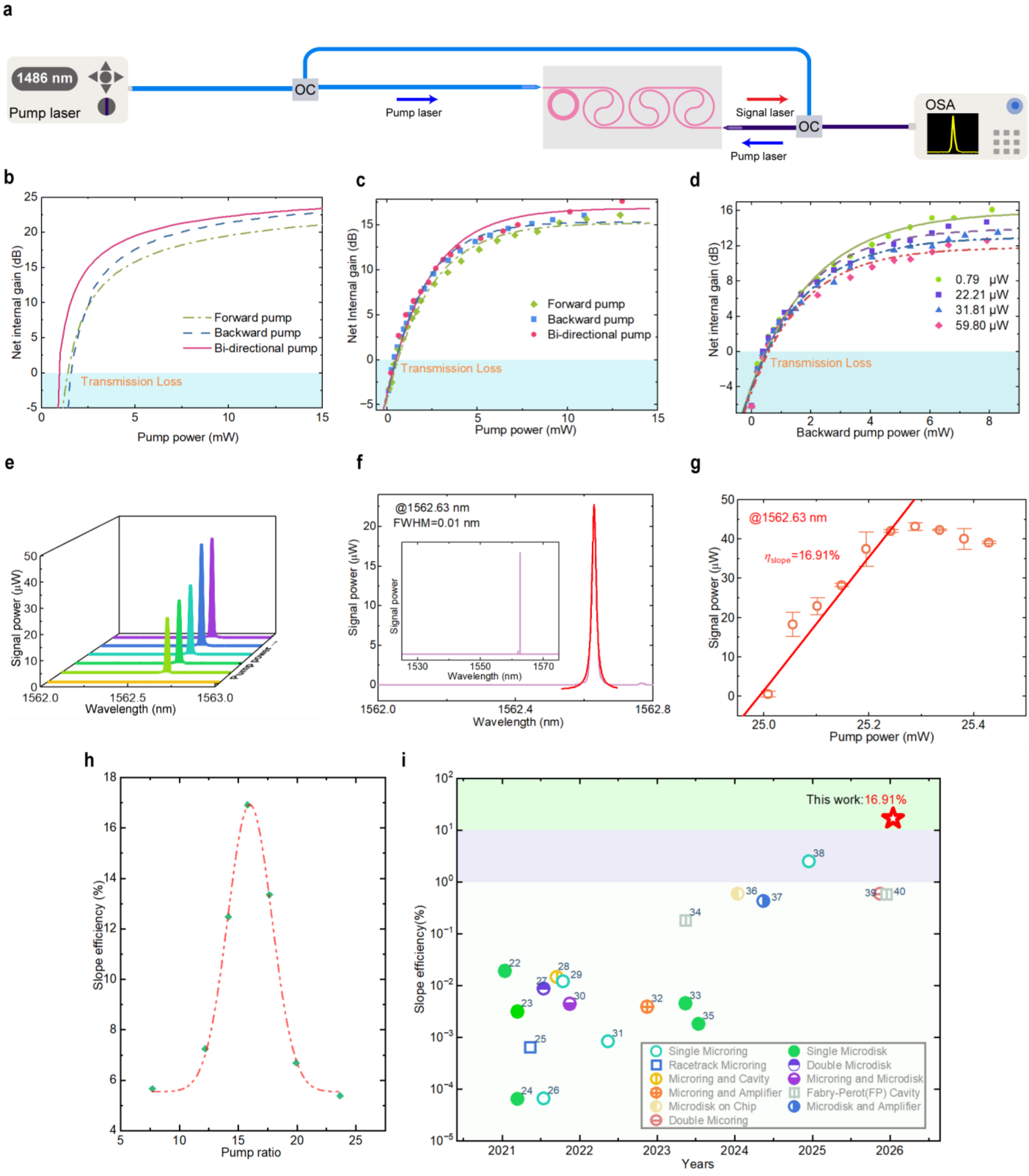

**Fig. 4. Characteristics of the integrated Er:LNOI laser under bidirectional pumping.** (a) Schematic of the optical setup. (b) Simulation results of the 3-turn waveguide amplifier under different pumping schemes. (c) Experimental results of the 3-turn waveguide amplifier under different pumping schemes. (d) Gain for different signal powers under backward pumping in a 3-turn waveguide amplifier. (e) On-chip lasing spectra corresponding to different on-chip pump powers. (f) Broadband and narrowband wavelength scans of the on-chip output laser spectrum obtained by an OSA. (g) On-chip output laser power versus pump power, demonstrating a slope efficiency of 16.91%. (h) Experimental results of laser slope efficiency under different pump ratios. (i) Slope efficiency versus publication year of Er:LNOI lasers, with our result highlighted by a red star.

## Discussion

Figure 4i summarizes recent progress in Er:LNOI lasers. The horizontal axis represents the publication year, the vertical axis denotes the slope efficiency, and different markers correspond to different device architectures. The number beside each marker indicates the corresponding reference, and the result of this work is highlighted by a red star. The evolution of slope efficiency can be broadly divided into two stages. From 2021 to 2023, the slope efficiency of Er:LNOI lasers increased from approximately $10^{-5}$% to $10^{-2}$%. Since 2023, most reported values have remained at the $10^{-1}$% level, except for one work that exceeded 1% using a microring resonator with a radius of 0.5 mm. This comparison suggests that a larger population of Er ions participating in stimulated emission is essential for achieving higher slope efficiency. Our results further show that increasing the Er doping concentration and enlarging the cavity size are constrained by crystal-growth quality and on-chip integration density. Therefore, it remains challenging to overcome the slope-efficiency bottleneck of Er LN lasers using a standalone microcavity resonator alone.

From Fig. 4i, it is evident that high-efficiency lasers in the 2023 to 2026 timeframe commonly adopt integration schemes with other auxiliary structures. We have noticed the high gain of Er:LNOI amplifiers, optimized the net gain of spiral amplifiers, monolithically integrated it with the microring laser, and achieved a slope efficiency as high as 3.45%. This result demonstrates that the amplifier plays a more critical role in the integrated laser performance at the current stage. The amplifier implemented in this work delivers a gain of 15.16 dB, while Er:LNOI amplifiers with gains exceeding 30 dB have been reported[49], indicating that the slope efficiency of integrated lasers can be further improved. On the other hand, bidirectional pumping has been proven effective for enhancing not only the gain of spiral amplifier but also the slope efficiency of integrated laser. For a standalone waveguide amplifier, the optimal bidirectional pump ratio is typically 1:1 to achieve maximum gain. But for the integrated laser, backward pump light that is not fully absorbed in the spiral waveguide will couple into the microring resonator and affect its output performance. Consequently, the backward pump intensity should be maintained significantly smaller than the forward pump intensity, resulting in an optimal bidirectional pump ratio substantially greater than 1.

In this work, we proposed a novel design strategy to enhance the slope efficiency of Er:LNOI laser: increasing the number of Er ions participating in luminescence, integrating the laser with an amplifier, and employing a bidirectional pumping scheme. Guided by these approaches, we grew high-quality and 1.0 mol% Er-doped LN crystals, integrated a microring resonator (200 μm radius, 40 μm width) with a 3-turn waveguide amplifier and adopted bidirectional pumping. The slope efficiency of Er:LNOI laser was increased from $5.86\times10^{-2}$% to 3.45% and then 16.91%, surpassing 10% for the first time on the LNOI platform. Our experimental results demonstrate that amplifier and bidirectional pumping are critical in the high output of integrated Er:LNOI laser. The design strategy presented here offers a new perspective and an implementable pathway for the future development of high efficiency LNOI lasers. This work provides a general route to boost lasing performance, filling a key gap in LN integrated photonics and enabling promising applications in compact optical communication and on-chip photonic signal processing systems.

## References

1. Hua, S. *et al.* An integrated large-scale photonic accelerator with ultralow latency. *Nature* **640**, 361-367 (2025).
2. Ambrogio, S. *et al.* An analog-AI chip for energy-efficient speech recognition and transcription. *Nature* **620**, 768-775 (2023).
3. Shalf J. The future of computing beyond Moore's Law. *Phil. Trans. R. Soc. A* **378**, 20190061 (2020).
4. Markov, I. L. Limits on fundamental limits to computation. *Nature* **512**, 147-154 (2014).
5. Elshaari, A. W., Pernice, W., Srinivasan, K., Benson, O. & Zwiller, V. Hybrid integrated quantum photonic circuits. *Nat. Photon.* **14**, 285-298 (2020).
6. Snigirev, V. *et al.* Ultrafast tunable lasers using lithium niobate integrated photonics. *Nature* **615**, 411-417 (2023).

7. Thomaschewski, M., Zenin, V. A., Wolff, C. & Bozhevolnyi, S. I. Plasmonic monolithic lithium niobate directional coupler switches. *Nat. Commun.* **11**, 748 (2020).
8. Wang, Z. *et al.* Ultracompact low-loss coupler between strip and slot waveguides. *Opt. Lett.* **34**, 1498-1500 (2009).
9. Han, J. *et al.* On-chip wavelength division multiplexing by angled multimode interferometer fabricated on erbium-doped thin film lithium niobate on insulator. *Nanophotonics* **13**, 2839-2846 (2024).
10. Renaud, D. *et al.* Sub-1 Volt and high-bandwidth visible to near-infrared electro-optic modulators. *Nat. Commun.* **14**, 1496 (2023).
11. Multani, K. K. S. *et al*. Integrated millimeter-wave cavity electro-optic transduction. *Nat. Commun.* **17**, 1166 (2026).
12. Cai, L. & Piazza, G. Low-loss chirped grating for vertical light coupling in lithium niobate on insulator. *J. Opt.* **21**, 065801 (2019).
13. Kang, S. et al. High-efficiency chirped grating couplers on lithium niobate on insulator. *Opt. Lett.* **45**, 6651-6654 (2020).
14. Dean, D. J. *et al.* Low-power integrated optical amplification through second-harmonic resonance. *Nature* **649**, 1159-1164 (2026).
15. Song, Y. *et al.* High-efficiency and broadband Kerr comb generation in normal-dispersion x-cut lithium niobate microresonators. *Sci. Adv.* **12**, eaeb5758 (2026).
16. Assumpcao, D. *et al.* A thin film lithium niobate near-infrared platform for multiplexing quantum nodes. *Nat. Commun.* **15**, 10459 (2024).
17. Wang, C. *et al.* Integrated lithium niobate electro-optic modulators operating at CMOS-compatible voltages. *Nature* **562**, 101-104 (2018).
18. Li, Q. *et al.* Ultra-broadband near- to mid-infrared electro-optic modulator on thin-film lithium niobate. *Nat. Commun.* **17**, 1138 (2026).
19. Wang, M. *et al.* Recent progresses on hybrid lithium niobate external cavity semiconductor lasers. *Materials* **17**, 4453 (2024).
20. Boes, A. *et al*. Lithium niobate photonics: Unlocking the electromagnetic spectrum. *Science* **379**, eabj4396 (2023).
21. Luo, Q., *et al.* Advances in lithium niobate thin-film lasers and amplifiers: a review. *Adv. Photon.* **5**, 034002 (2023).
22. Wang, Z. *et al.* On-chip tunable microdisk laser fabricated on $Er^{3+}$-doped lithium niobate on insulator. *Opt. Lett.* **46**, 380-383 (2021).
23. Liu, Y. *et al.* On-chip erbium-doped lithium niobate microcavity laser. *Sci. China Phys. Mech. Astron.* **64**, 234262 (2021).
24. Luo, Q. *et al.* Microdisk lasers on an erbium-doped lithium-niobate chip. *Sci. China Phys. Mech. Astron.* **64**, 234263 (2021).
25. Yin, D. *et al.* Electro-optically tunable microring laser monolithically integrated on lithium niobate on insulator. *Opt. Lett.* **46**, 2127-2130 (2021).
26. Luo, Q. *et al.* On-chip erbium-doped lithium niobate microring lasers. *Opt. Lett.* **46**, 3275-3278 (2021).
27. Gao, R. *et al.* On-chip ultra-narrow-linewidth single-mode microlaser on lithium niobate on insulator. *Opt. Lett.* **46**, 3131-3134 (2021).
28. Xiao, Z. *et al*. Single-frequency integrated laser on erbium-doped lithium niobate on insulator. *Opt. Lett.* **46**, 4128-4131 (2021).
29. Li, T. *et al.* A single-frequency single-resonator laser on erbium-doped lithium niobate on insulator. *APL Photonics* **6**, 101301 (2021).
30. Liu, X. *et al.* Tunable single-mode laser on thin film lithium niobate. *Opt. Lett.* **46**, 5505-5508 (2021).
31. Liang, Y. *et al.* Monolithic single-frequency microring laser on an erbium-doped thin film lithium niobate fabricated by a photolithography assisted chemo-mechanical etching. *Opt. Continuum* **1**, 1193-1201 (2022).
32. Zhou, J. *et al.* Laser diode-pumped compact hybrid lithium niobate microring laser. *Opt. Lett.* **47**, 5599-5601 (2022).
33. Gao, R. et al. Electro-optically tunable low phase-noise microwave synthesizer in an active lithium niobate microdisk. *Laser Photonics Rev.* **17**, 2200903 (2023).
34. Yu, S. *et al.* On-chip single-mode thin-film lithium niobate Fabry–Perot resonator laser based on Sagnac loop reflectors.

*Opt. Lett*. **48**, 2660-2663 (2023).
35. Luo, Q. *et al*. On-chip erbium–ytterbium-co-doped lithium niobate microdisk laser with an ultralow threshold. *Opt. Lett.* **48**, 3447-3450 (2023).
36. Guan, J. *et al*. Monolithically integrated narrow-bandwidth disk laser on thin-film lithium niobate. *Opt. Laser Technol.* **168**, 109908 (2024).
37. Wu, J. *et al.* Efficient integrated amplifier-assisted laser on erbium-doped lithium niobate. *ACS Photonics* **11**, 2114-2122 (2024).
38. Sun, C. *et al*. High-efficiency single-mode erbium-doped lithium niobate microring laser with milliwatt output power. *Opt. Lett.* **49**, 6996-6999 (2024).
39. Hou, Q. *et al.* Monolithic tunable single-frequency microlaser on erbium-doped lithium niobate on insulator. *Opt. Laser Technol*. **190**, 113242 (2025).
40. Hou, G. *et al*. On-chip laser emission from a 2-μm-thick $Er^{3+}$-doped lithium niobate on insulator. *Opt. Express* **33**, 53075-53083 (2025).
41. Kalichevsky-Dong, M. T. *et al*. Diffraction-limited 308 W at ~980 nm from a monolithic ytterbium fiber laser. *Opt. Fiber Technol.* **95**, 104396 (2025).
42. Meng, X. *et al*. A 2 kW fully open cavity random Raman fiber laser employing tapered fiber. *High Power Laser Sci. Eng.* **13**, e89 (2025).
43. Kamrádek, M. et al. Nanoparticle doping as a way to enhance holmium fiber lasers efficiency. *Opt. Commun.* **575**, 131290 (2025).
44. Luo, Q. *et al.* On-chip erbium-doped lithium niobate waveguide amplifiers. *Chin. Opt. Lett.* **19**, 060008 (2021).
45. Wang, Y. et al. Erbium-doped lithium niobate on insulator waveguide amplifier with ultra-high internal net gain of 38 dB. In CLEO 2024, Technical Digest Series, paper ATu4M.3 (Optica Publishing Group, 2024); https://doi.org/10.1364/CLEO_AT.2024.ATu4M.3.
46. Wei, Z. *et al.* Erbium-doped lithium niobate waveguide amplifier enhanced by an inverse-designed on-chip reflector. *Opt. Lett*. **50**, 3624-3627 (2025).
47. Zhang, Y. *et al.* Efficient self-amplification in monolithically integrated laser on Er/Yb co-doped lithium niobate on insulator. *Laser Photonics Rev.* e71269 (2026).
48. Zhang, Z. *et al.* The doping concentration optimization of Er-doped $LiNbO_3$ crystals for LNOI lasers and amplifiers. *Cryst. Growth Des*. **24**, 6838-6844 (2024).
49. Wang, Y. *et al.* Unifying optical gain and electro-optical dynamics in Er-doped thin-film lithium niobate platform. *Nat. Commun.* **16**, 10462 (2025).
50. Luo, L.-W. *et al*. High quality factor etchless silicon photonic ring resonators. *Opt. Express*. **19**, 6284-6289 (2011).